\documentclass{pr-imfp00}
\def\beq{\begin{equation}\nonumber}%
\def\eeq{\end{equation}\nonumber}%
\def\bea{\begin{eqnarray}}%
\def\eea{\end{eqnarray}}%
\begin{document}

\title{Experimentation and Physics at a Future Electron-Positron Linear Collider}
\footnotetext[1]{Invited talk given at the XXVIII
International Meeting on Fundamental Physics, Sanl\'ucar de Barrameda, C\'adiz, Spain,
14-18 February 2000}

\author{Martin Pohl}

\address{DPNC, Universit\'e de Gen\`eve, Switzerland\\and\\
EHEF, Katholieke Universiteit Nijmegen, The Netherlands\\
E-mail: Martin.Pohl@cern.ch}


\maketitle

\abstracts{ I summarise the physics opportunities and experimental
challenges at future Linear  Colliders, using material from the recent
ECFA/DESY workshop on  the subject as well as contributions to the
series of worldwide studies. For reasons of economy,  the discussion
is restricted to the European Tesla project and to its
electron-positron mode only.}

\section{Yet Another Accelerator?}\label{sec:intro}

Experimentation at the current generation of high energy colliders, LEP at CERN and
the Tevatron at FNAL, is not yet completed. The next generation hadron collider LHC at CERN
is on track but far from finished.
One may thus wonder why now is the right time to discuss future accelerator projects.
The reason is that the recent past shows that a new generation of accelerators
takes of order 10 years to prepare. It takes another $\sim\!10$ years to build and 
commission the machine and its experiments. Consequently, the preparations for
a successor to LEP started in the early 1990. And it is now the time to design 
the next lepton machine and its integrated experiments -- and get them approved --
to answer the physics questions of 2010. 

It is obvious to everyone that the current generation of machines and experiments
surpassed expectations by a large margin. LEP and SLC established the electroweak
sector of the Standard Model at the loop level, with such precision that even
the mass of inaccessible particles is predicted from radiative corrections with
respectable precision\cite{Mnich00}. The Tevatron experiments discovered the top quark 
at the mass required for radiative corrections and thus
completed the periodic system of matter required by the Standard Model.
Nevertheless, as good an effective theory as the Standard Model may be, it still
remains an effective theory with many unanswered questions. We list only a few:
\begin{itemize}
\item Current data do not exclude the existence of a Higgs boson,
provided that it is rather light. But we still miss experimental evidence for this 
mechanism of mass generation proposed by the Standard Model. 
\item Gravity is completely missing from the current picture based on gauge theories.
\item A rationale for the values of couplings and their potential unification
is not provided by the Standard Model. In particular the structure and values of the 
CKM\cite{Cabibbo63,Kobayashi73} and MNS\cite{Maki62} matrix elements remain a complete mystery.
\item The number and hierarchy of fermion generations remains unexplained.
\item It is not known why matter apparently consists of fermions only and why forces 
are transmitted by vector bosons. Supersymmetry proposes to lift this
asymmetry. 
\end{itemize}
LHC will clearly greatly contribute to approaching answers to these questions. If
a light Higgs particle exists, it will be discovered at LHC. 
If it does not exist, LHC might reveal interactions among electroweak
gauge bosons, resonances at the TeV scale or other mechanisms to cut off 
electroweak divergences. Structures of space-time responsible for gravitation might
be discovered. And last but not least,
a large part of the rich spectrum of supersymmetric particles will be
observed at LHC if nature chooses to be supersymmetric. Physics after LHC will thus
be less focused on the discovery of new physics but on its proper understanding.
One can foresee that this effort will focus on three main themes:
\begin{itemize}
\item refined knowledge of model parameters, especially the properties
of the top quark and the W boson as well as those of supersymmetric particles;
\item studies of the Higgs boson -- if it exists -- to establish the
Higgs-like nature of the particle, to measure its couplings to matter and its
self-coupling and to distinguish between the standard Higgs and its supersymmetric
alternatives;
\item understanding of alternative mechanisms -- if the Higgs mechanism is not 
realised -- in the study of new states in boson scattering or in 
the search for a substructure of matter.
\end{itemize}
In a sense, LHC will thus point out the way towards a new effective theory replacing the
Standard Model. And a second generation machine will be required to establish
this new effective theory, much in the way that LEP established the Standard Model
itself. I believe that such a machine should be an electron-positron collider.
While not so much fit for exploratory studies into a
completely unknown energy domain, their fixed beam energy, fully available for 
annihilation reactions and providing constrained final state kinematics, make them
ideal machines for dedicated studies in well defined energy ranges. 

Electron-positron cross sections in the interesting energy range, from
top threshold to 1 TeV, are much smaller than cross sections at LEP. One thus
needs a machine that exceeds current $\mathrm{e^+ e^-}$ collider luminosities
by two orders of magnitude. 
Synchrotron radiation power losses prevent us from extrapolating the successful
technology of circular electron-positron colliders to higher energies. The
power loss, $P_B$, on a circular orbit of radius $R$ is proportional to 
\beq P_B \sim \frac{E^3_b}{m^3 R}
\eeq
with the particle energy $E$ and its mass $m$. LEP, with a circumference of 27 km, 
will thus be the largest circular $\mathrm{e^+ e^-}$ collider ever to be built.
There are two ways to efficiently fight radiation energy loss in a lepton collider.
One is to use heavier leptons as is done in muon collider 
projects\cite{Ankenbrandt99}. The other one is to take the radius to infinity,
which is the idea behind Linear Colliders, pioneered with the SLC.
In addition, Linear Colliders feature potentially high beam polarisation, crucial
for the measurement of spin dependent amplitudes and the suppression of background.
I thus believe that a Linear electron-positron Collider is the right choice for
a first generation machine directly after LHC.

This report is based on material from the {\em Second Joint ECFA/DESY Study on Physics 
and Detectors for a Linear Electron-Positron Collider}\cite{EcfaDesy98}
as well as the {\em Worldwide Study on the Physics and Detectors for Future
Linear $\mathrm{e^+ e^-}$ Colliders}\cite{World98}. Most of the material comes
from the Tesla Conceptual Resign Report\cite{Brinkmann97,Accomando98} (CDR), 
the Sitges workshop of the worldwide study\cite{Sitges99} and the Obernai workshop 
of the ECFA/DESY study\cite{Obernai99}.

\section{The Tesla Machine}

The figures of merit that characterise any machine for the experimenter are the 
center of mass energy, $\sqrt{s}$, and the luminosity, ${\cal L}$.
Superconducting Linear Colliders present advantages over normal conducting ones.
I will concentrate on only two aspects: the high power 
conversion efficiency and the comparatively relaxed alignment tolerances.

The luminosity of a collider can be written as\cite{Brinkmann97a}
\beq
{\cal L} = \frac{n_b N_e^2 f}{4 \pi \sigma_x \sigma_y} H_D
\eeq
where $n_b$ is the number of bunches/pulse, 
$N_e$ the number of $\mathrm{e^\pm}$ per bunch,
$f$ the pulse repetition frequency and
$\sigma_{x,y,z}$ are the horizontal, vertical and longitudinal beam sizes at 
the interaction point. The disruption enhancement factor, 
$H_D$, is due to the pinch effect that focuses one bunch in the field of the other;
it has typical values of $\simeq 1.5$. 

\begin{figure}[htbp]
\begin{center}
\includegraphics*[width=0.5\linewidth,clip=]{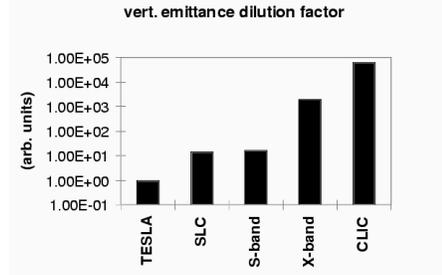} 
\end{center}
\caption{Comparison of the expected vertical emittance dilution factor $F$ for
the Tesla low frequency, superconducting Linear Collider compared to high 
frequency normal conducting ones. \label{fig:epsilon}}
\end{figure}

The luminosity can also be given in terms of the average beam power, $P_b$,
at a given center of mass energy, $\sqrt{s}$,
\beq
P_b = n_b N_e f \sqrt{s} = \eta P_{AC}
\eeq
which is the product of the AC power consumption, $P_{AC}$, and the power
conversion efficiency, $\eta$. This efficiency can be as high as 16\% for a 
superconducting Linear Collider, which is the first of the major advantages of this 
technology. The remaining parameters of the luminosity can be expressed in terms 
of the fraction, $\delta_E$,  of beam energy lost by collective beamstrahlung
\beq
\delta_E = \frac{\Delta E_b}{E_b} 
\sim \frac{r_e^3 N_e \gamma}{\sigma_z (\sigma_x+\sigma_y)^2}
\sim \frac{r_e^3 N_e \gamma}{\sigma_z \sigma_x^2}
\eeq
The rightmost proportionality is reached in the limit of flat beams
$\sigma_x \gg \sigma_y$, which allows to maximise luminosity without intolerable
beamstrahlung losses by reducing the vertical beam size $\sigma_y$. The beam phase
space at the interaction point is given by the emittance, $\epsilon$, and the $\beta$ 
function
\beq
 \sigma_y = \sqrt{\epsilon_y \beta_y}
\eeq
where a $\beta_y \simeq \sigma_z$ corresponds to a  practical limit. The 
luminosity in these terms thus becomes
\beq
{\cal L} \sim \frac{\eta P_{AC}}{\sqrt{s}} 
\sqrt{\frac{\delta_E}{\epsilon_y}} H_D
\eeq
inversely proportional to the square root of the emittance.
The second major advantage of superconducting technology is that it allows for low 
emittance dilution by short range wake fields. This dilution depends on the 
r.m.s.~lattice misalignment, $\delta\!y^2$,
\beq
\frac{\Delta \epsilon}{\epsilon} \sim F\,\bar{\beta}\,\delta\!y^2
\eeq
with a dilution factor, $F$, which is extremely dependent on the radiofrequency,
$f_{RF}$,
\beq
F \sim \frac{N_e^2 \sigma_z f_{RF}^6}{\epsilon_y}
\eeq
Fig.~\ref{fig:epsilon} shows a comparison between the 
expected dilution factors for the vertical emittance of Tesla as compared to 
high frequency normal conducting machines which are under study in Japan\cite{Jlc99}
and in the USA\cite{Nlc99} and which aim at similar performance.
A low frequency, typical for superconducting accelerators, is thus very beneficial
in that it reduces the requirements for alignment tolerances, yet preserving
high luminosity. 

It is unrealistic to expect that accelerating gradients in a superconducting cavity
were to rise above some 40 MV/m. Surface defects that even the most careful conditioning
cannot prevent, limit realistic gradients to such values. It is therefore clear that a 
superconducting Linear Collider like Tesla will not enter into the TeV energy domain.
Beyond that frontier, novel accelerating techniques like the two-beam accelerator CLIC
will have to come into play. However, a machine with up to 1 TeV energy indeed covers most
of the LHC's discovery range and is thus entirely appropriate as a first generation
machine for the era after LHC.

\begin{figure}[htbp]
\begin{center}
\includegraphics*[height=0.85\textheight,bb=0 0 340 735]{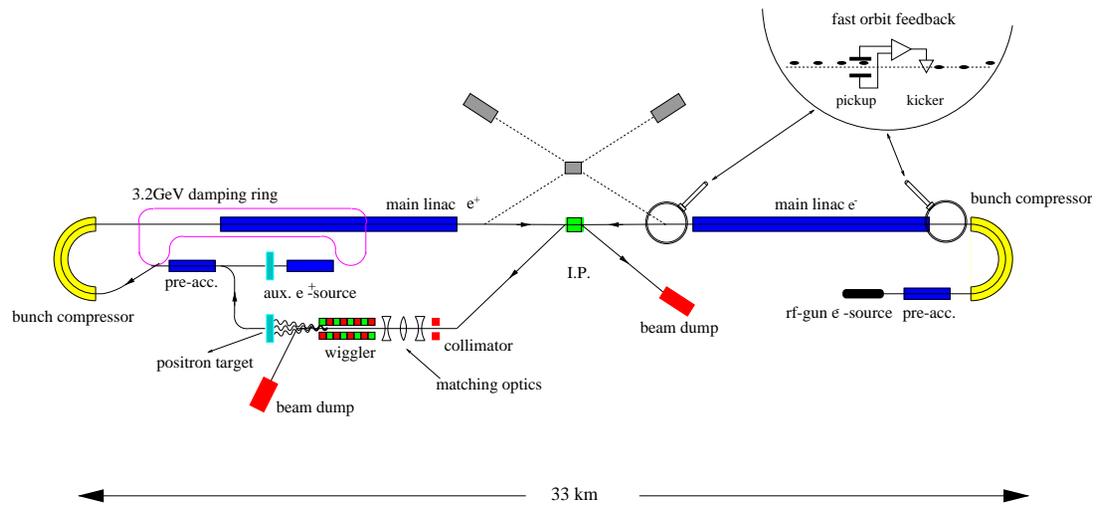}
\end{center}
\caption{Layout of the Tesla machine complex (not to scale) showing the principle
machine components.\label{fig:layout}}
\end{figure}

I will limit further discussions to the proposed Tesla machine.
Fig.~\ref{fig:layout} shows the layout of the machine complex. Its major
parameters are summarised in Tab.~\ref{tab:tesla} and compared to the 
performance of LEP. Mind that the contents of this table is my own interpretation of 
numbers from various sources and not an official Tesla design specification. It
clearly includes some amount of wishful thinking of an experimenter {\em in spe}.

\begin{table}[htbp]
\caption{Comparison of machine performance parameters between what has actually
been achieved at LEP 200 and what the eager experimenter might hope for
at Tesla. This is not a compilation of official Tesla design goals.\label{tab:tesla}}
\begin{center}
\footnotesize
\begin{tabular}{|l|c|cc|}
\hline
                         & LEP 200      & \multicolumn{2}{c|}{TESLA}           \\ \hline
type                     & Storage Ring & \multicolumn{2}{c|}{Linear Collider} \\
maximum energy           & 200 GeV      & 500 GeV     & 800 GeV                \\
total length             & 26.7 km      & \multicolumn{2}{c|}{33km}            \\
accelerating gradient    & 6 MV/m       & 25 MV/m     & 40 MV/m                \\
maximum bunches          & 8            & 1410        & 4028                   \\
beam size at IP [$\mu$m] & 150$\times$10 & 0.85$\times$0.02 & 0.34$\times$0.02 \\
luminosity [cm$^{-2}$s$^{-1}$] 
                       & 10$^{32}$ & 1$\times$10$^{34}$ & 5$\times$10$^{34}$ \\ \hline
\end{tabular}
\end{center}
\end{table}

\begin{figure}[htbp]
\begin{center}
\raisebox{2mm}{\includegraphics*[width=0.49\linewidth]{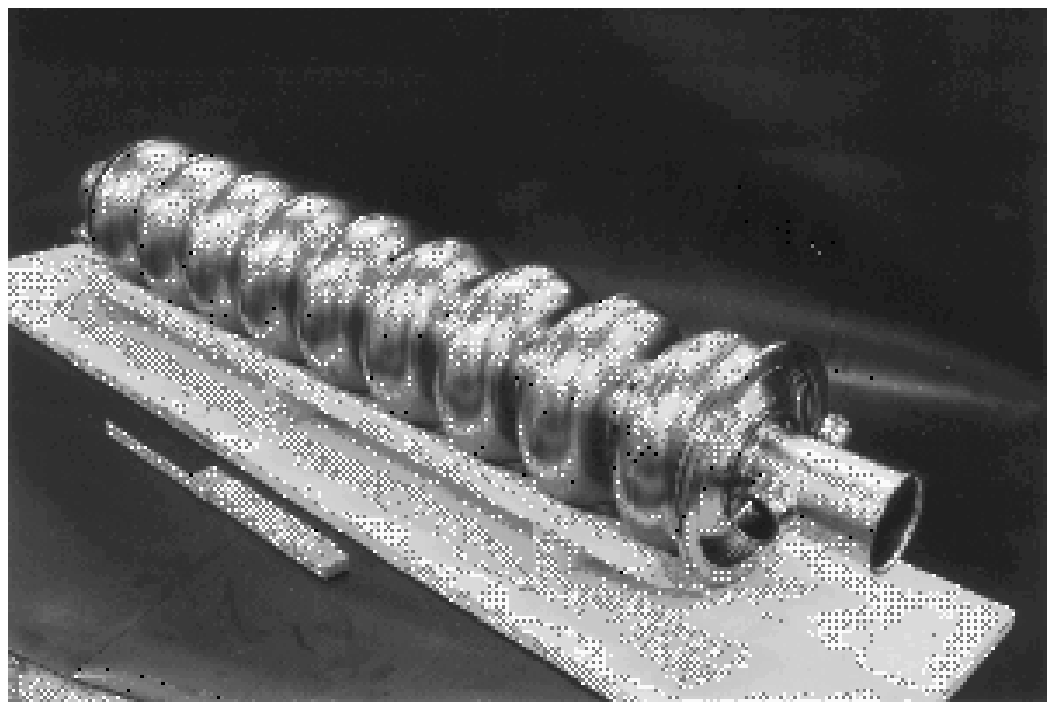}}
\includegraphics*[width=0.49\linewidth]{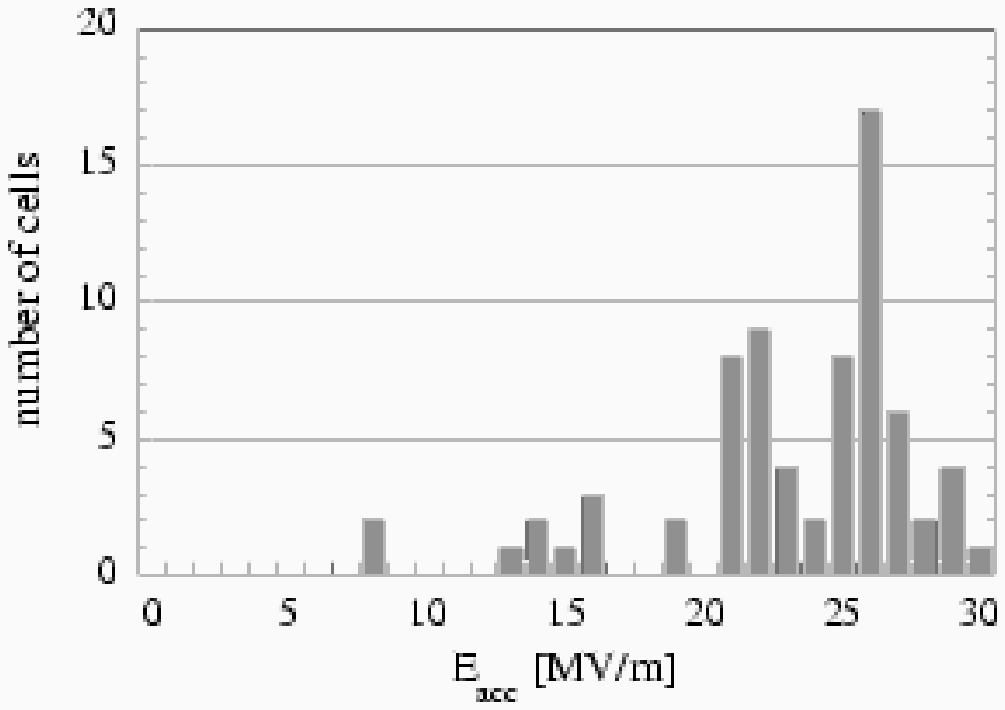} 
\end{center}
\caption{Left: Nine cell niobium cavity for Tesla. Right: Statistics of accelerating 
gradients in early cavity production. Performance
exceeding specifications are routinely reached by industry after a short learning 
phase.\label{fig:cavity}}
\end{figure}

One of the major challenges in reaching these ambitious goals is the accelerating
gradient in excess of 25 MV/m required to keep the tunnel length inside reasonable 
limits. Fig.~\ref{fig:cavity} shows a nine-cell niobium superconducting cavity 
of which Tesla will need about twenty thousand. These are now produced by several 
European companies and tested at the DESY Tesla Test Facility. In recent
production, both the quality factor and the maximum gradient have consistently 
exceeded the requirements. As an example, Fig.~\ref{fig:cavity} shows the statistics
of accelerating gradients. The more experience in production and post-production 
conditioning is gained, the more cavities exceed the required minimum gradients.

Another challenge is in the final focus, required to produce the
tiny beam sizes at the interaction point. 
Final focus test facilities are available at SLAC and DESY.
The beam delivery system with collimation, tuning section and final transformer will
be several kilometers long, a substantial fraction of the total collider length. 
The final transformer will demagnify the beam by a factor of 15 in the vertical
and 50 in the horizontal plane. Such large focusing forces inevitably create
a large amount of synchrotron radiation close to the interaction point which 
have to be shielded by a sophisticated system of masks. Their unshieldable
component as well as the closest mask has to be accommodated by a 
forward-backward hole in the detector (see Fig.~\ref{fig:vertex}).

The whole Tesla complex will be housed in a single 5m diameter tunnel, with minimal
environmental impact, including
the main linac and the damping ring as well as all services.

\section{A Tesla Detector}

\begin{figure}[htbp]
\begin{center}
\includegraphics*[width=0.5\linewidth]{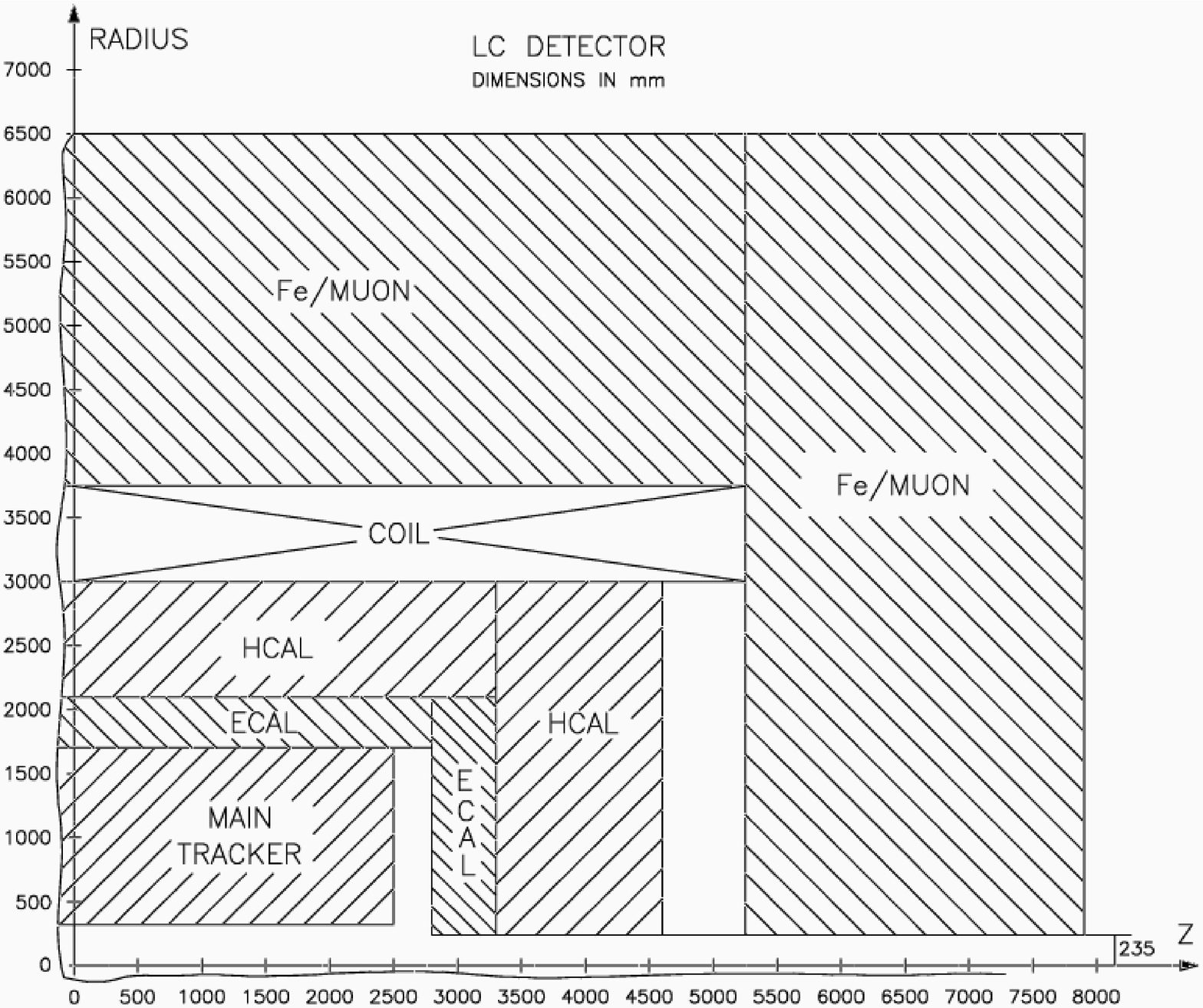}
\end{center}
\caption{Sketch of the Tesla detector layout as described in the TDR.\label{fig:detlayout}}
\end{figure}

As will be shown below, physics at a Linear Collider requires a detector with truly
excellent performance. Physics requirements focus on three main aspects:
\begin{itemize}
\item {\em Hadronic jets:} an excellent resolution for charged particles
as well as high granularity calorimetry are required to obtain the necessary
resolution on jet energy and direction as well as jet-jet mass. Identification of
bottom and charm jets is a must, thus excellent vertexing is required.
\item {\em Leptons:} electron-hadron distinction requires excellent tracking
resolution as well as electromagnetic energy resolution. Again, fine grain calorimetry 
must allow to tag electrons close to a jet. Obviously, muon identification is a must.
\item {\em Missing energy and momentum:} In order to take advantage of the constrained
final state kinematics, one needs redundancy in the tracker and calorimeter subsystems. 
The energy flow methodology must be used to fully take advantage of this redundancy.
The detector has to be hermetic down to the mask shielding synchrotron radiation.
\end{itemize}
A conservative model detector that fulfils these basic requirements has been 
described in the Tesla Conceptual Design Report\cite{Brinkmann97}. Its layout is shown in
Fig.~\ref{fig:detlayout}. I will summarise its 
main design features in the following. I will also point out less conservative alternatives
for subsystems that are being discussed presently.

It is tempting to scale one of the LEP detectors to obtain a viable model for a
Linear Collider detector at energies two to four times higher. And indeed the detector
described in the Tesla CDR conceptually resembles the Aleph detector rather closely.
Nevertheless, the smaller beam pipe allows for instrumentation down to smaller radius,
with finer grain detectors of higher resolution, as pioneered by SLD. Although the average 
particle energy varies little with the center of mass energies in hadron production, the 
maximum momenta in low multiplicity leptonic final states are of the order of the beam energy.
Thus, better point resolutions, more measured points on the trajectory and higher
magnetic fields are required for the tracking device. And the higher maximum electron,
photon and hadron energies point towards thicker electromagnetic and hadronic calorimeters.
The subdetectors of the Tesla detector study have been dimensioned accordingly, as
shown in Fig.~\ref{fig:detlayout}. 

The inner detector, with the tracker and the full calorimeters, is enclosed
in a superconducting solenoidal magnet of 6m inner diameter and more than 10m length. The CDR
study foresaw 3T field strength, with a total stored energy of 1300 MJ. Since then,
studies for LHC magnets -- like the CMS\cite{Kircher98} one --
have shown that higher stored energies are
feasible. Optimisation towards a 4T magnet, storing 2500 MJ, is thus in progress. This will 
have implications on the machine itself, and influence experimental conditions like the 
number of spiralling tracks, tracker occupancy and resolution.
The magnet is surrounded by an almost 3m thick iron yoke of about 70\% filling factor. 
The remaining 30\% of the volume consist of instrumentation for muon detection. 
A simple large surface detector, like resistive plate chambers\cite{Benvenuti90}
arranged in multiple layers, will suffice to provide optimum protection against 
hadron misidentification.

\begin{figure}[htbp]
\begin{center}
\includegraphics*[width=\linewidth]{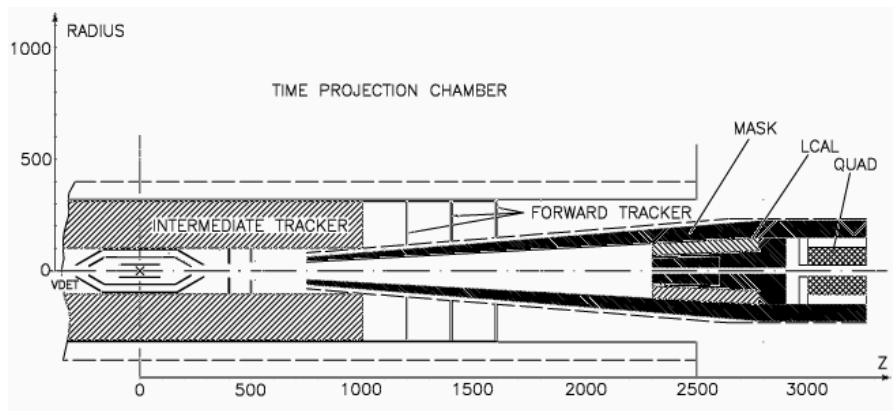}
\end{center}
\caption{Layout of a pixel vertex detector surrounding the interaction point, supplemented
by forward disks surrounding the mask.\label{fig:vertex}}
\end{figure}

As has been demonstrated at SLC, Linear Collider beam pipes at the interaction point can 
have a radius as small as 2cm, thus leading to a first layer of tracking detector at
about 2.5cm from the beam. The use of strip detectors at such a small radius is excluded 
because of track overlap: a fraction of overlapping hits of order thirty percent would 
results for realistic strip lengths. Thus, pixel detectors\cite{Heyne89}
are a must. At around 500$\mu$m double hit resolution, they suffer no more
than 5\% of overlapping hits. Fig.~\ref{fig:vertex} shows a possible layout consisting
of three cylindrical layers with cone shaped end caps for the outer layers. At small
angle, disk shaped forward layers complete the set-up. As far as the sensor itself is 
concerned, the CCD\cite{Damerell94} technology pioneered at SLD
is clearly a prime candidate. Alternatively, active pixel detectors are very interesting. 
Beyond the bump-bonded technology used by e.g.~Delphi\cite{Becks97}, 
monolithic active pixel sensors\cite{Caccia99} are most promising.

\begin{figure}[htbp]
\begin{center}
\includegraphics*[width=.5\linewidth]{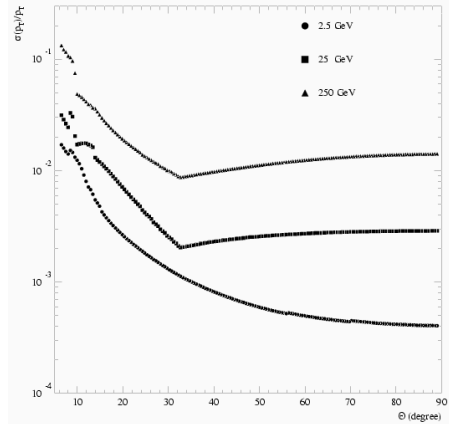} 
\end{center}
\caption{Resolution in the transverse momentum $p_t$ as a function of angle for the Tesla
tracking detector at three representative momenta.\label{fig:pres}}
\end{figure}

For the main tracker, there is hardly any serious alternative to a 
Time Projection Chamber (TPC)\cite{Decamp90,Buskulic95}, since no other detector
features such a high density of three-dimensional hits with good resolution. In addition,
a TPC provides $dE/dx$ measurements for particle identification. Octagonal and circular
cross sections are under study. Innovative improvements
to cluster detection at the end plates, using the GEM\cite{Sauli97} device, are under
development. With its large radius,
the projected Tesla TPC, together with the vertex detector, would reach a superb 
momentum resolution as shown in Fig.~\ref{fig:pres}.

\begin{figure}[htbp]
\begin{center}
\includegraphics*[width=0.69\linewidth]{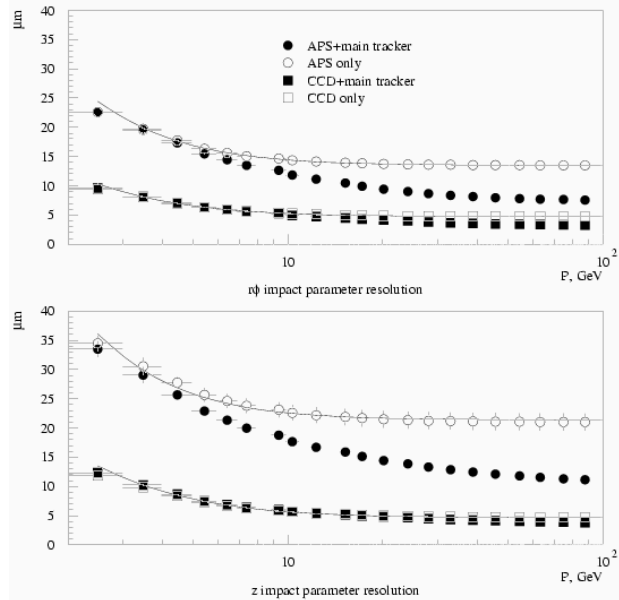}
\end{center}
\caption{Resolution in the transverse and longitudinal impact parameter for the projected 
Tesla main tracker and two representative vertex detector performances.\label{fig:dres}}
\end{figure}

Likewise the vertex detector would outperform currently available detectors in a impressive
way. Impact parameter resolutions as shown in Fig.~\ref{fig:dres} are expected,
with constant terms of order 5$\mu$m and multiple scattering terms of order 30$\mu$m$/p$
(with $p$ in GeV) for the example of the CCD option. Such performances are indeed
required for the measurement of the Higgs Yukawa couplings as shown in Sect.~\ref{sec:higgs}.

The Tesla CDR proposes a conservative fine grain calorimeter of ``shashlik'' readout,
inspired by similar DELPHI\cite{Benvenuti93} and H1\cite{Hartouni95} calorimeters.
In this scheme, a metal/scintillator sandwich is read out radially by wave length 
shifting fibers crossing absorber plates. The
Tesla baseline calorimeter has a physical tower structure. The towers have a granularity
of $1^\circ \times 1^\circ$ in the electromagnetic section, where the absorber is 
lead and the longitudinal sampling is fine. In the hadronic section, made of copper and
scintillator, the towers are $2^\circ \times 2^\circ$ wide and the sampling is coarser.
All towers almost point to the interaction point, with a small angular deviation to
avoid radial cracks in the acceptance. The Tesla CDR quoted rather optimistic
performance projections, with a resolution for electromagnetic showers of order
$\sigma/E \simeq 0.10/\sqrt{E} \oplus 0.01$ and jet energy resolutions of
$\sigma/E \simeq 0.57/\sqrt{E} \oplus 0.01$ (with $E$ in GeV). These estimates need to be 
substantiated by further simulation studies including more detail on dead material.

Large shashlik type calorimeters are difficult
to build since transverse forces tend to stress or even shear off the readout fibers.
Therefore, and to optimise space and resolution, other readout schemes, like the 
tile calorimeter\cite{Henriques95} with transverse readout are under study.
Also radically different and interesting approaches like a silicon-tungsten
sampling calorimeter are envisaged for the electromagnetic section. This 
technology has so far been restricted to small luminometers\cite{Anderson94,Bederede95}
but larger scale applications have very attractive features\cite{Videau99,Brient99} indeed: 
the resolution is excellent, very fine granularity can be realised in a three-dimensional
readout and the set-up is extremely compact. With these technologies, simpler octagonal
physical structures can be designed, with towers implemented only in readout.

\begin{figure}[htbp]
\begin{center}
\includegraphics*[width=0.55\linewidth]{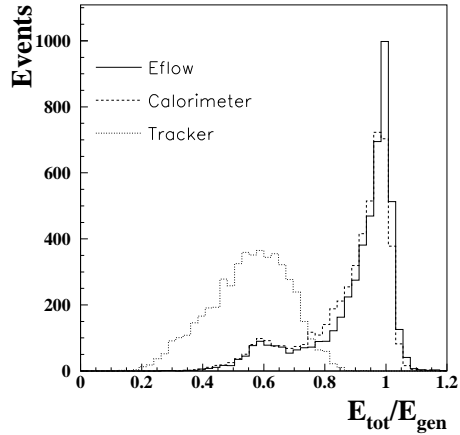}
\end{center}
\caption{Resolution for the total event energy in hadronic events at 800 GeV
as determined from energy flow measurement (solid line), as compared to
calorimetric measurements only (dashed line) and charged track momenta only (dotted line).
\label{fig:eflow}}
\end{figure}

It is of course not only the subdetector performance that determines performance for
physics, but also how they work together, complement each other and create
synergy. Tracker and calorimeter measurements of charged particle momentum and energy
have to be combined to provide optimum resolution from low to high
momenta. Neutral electromagnetic and hadronic energy deposits have to identified
in the calorimeters and separated off charged energy to avoid double counting. The
energy flow approach\cite{Pohl99} provides an optimum framework to do this. There, tracker 
and calorimeter information is combined to provide a system's resolution that is better
than the subdetector performance as demonstrated in Fig.~\ref{fig:eflow}.

\section{Top Physics}\label{sec:top}

\begin{figure}[htbp]
\begin{center}
\includegraphics*[width=0.48\linewidth]{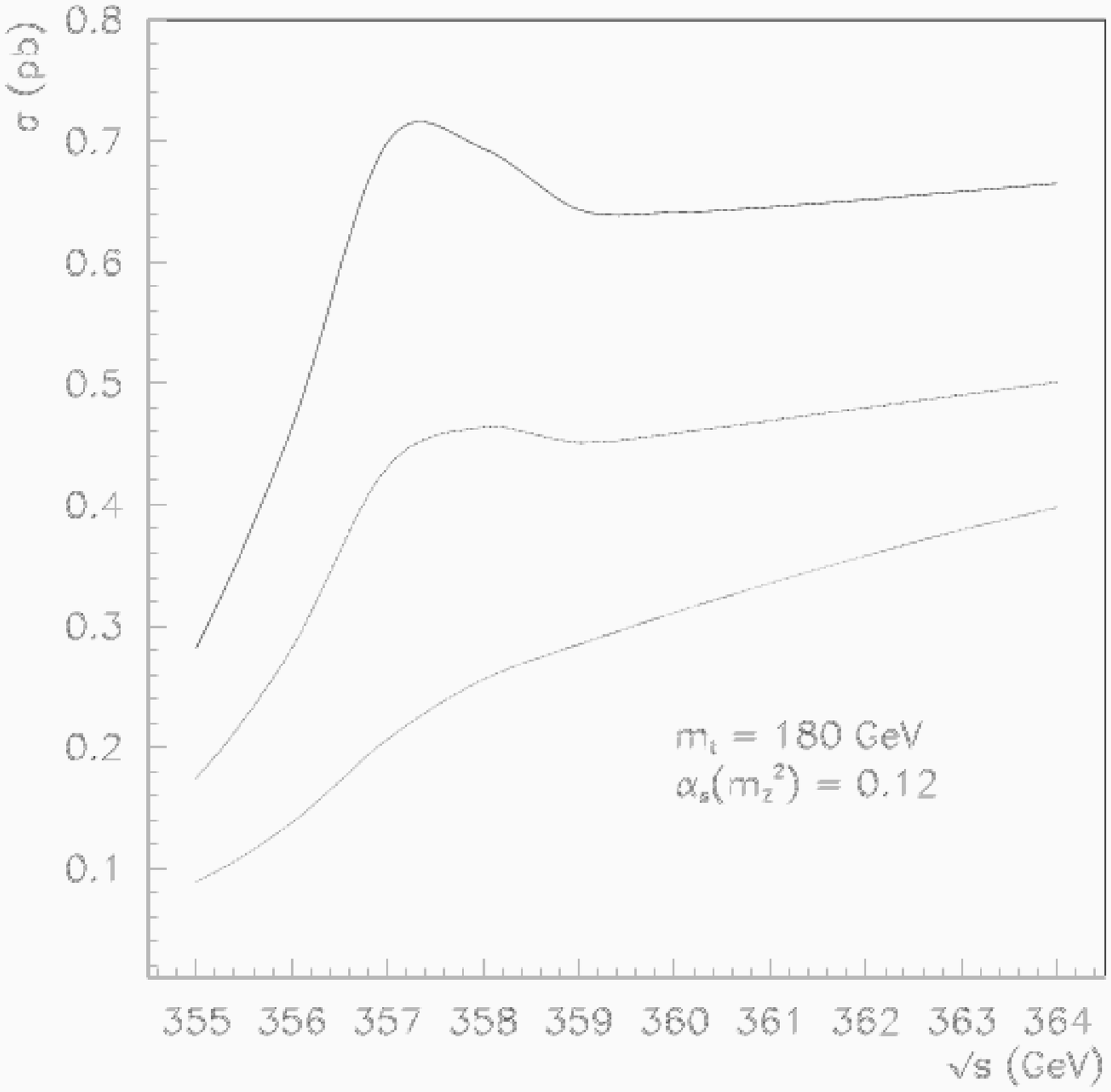} 
\includegraphics*[width=0.50\linewidth]{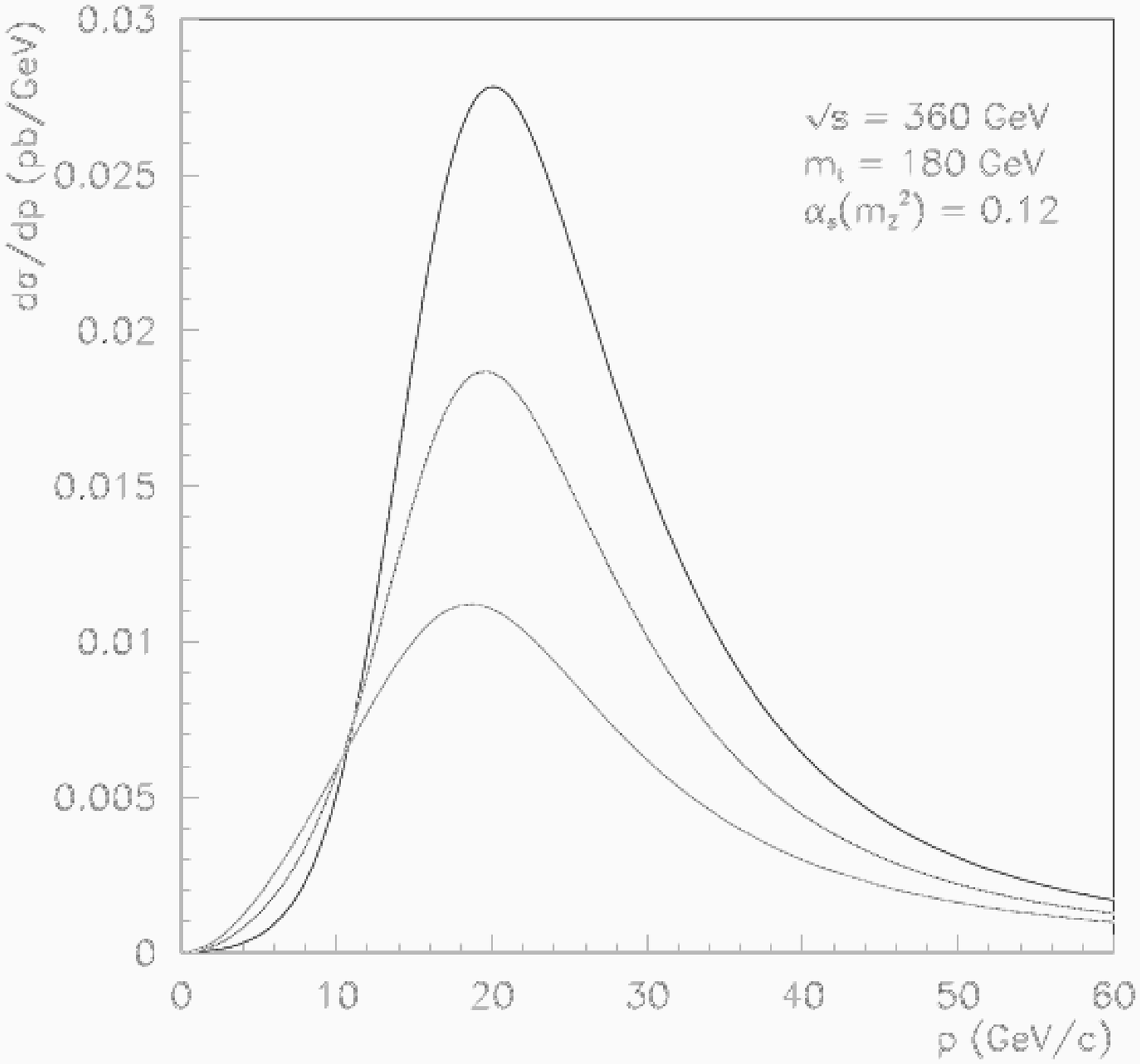} 
\end{center}
\caption{Top production cross section (left) and momentum distribution (right) 
at Born level (solid line), with initial state radiation (dashed line) and with beamstrahlung 
(dotted line).\label{fig:top}}
\end{figure}

A precision measurement of the top quark mass, $m_{\mathrm{t}}$, is required to over-determine 
electroweak parameters. The most promising method to reach sub-GeV accuracy is to determine 
the position of the top production threshold\cite{Comas96}.
Unlike the lighter quarks, the top quark decays too quickly to form vector meson resonances
in $\mathrm{e^+ e^- \rightarrow t \bar{t}}$ below threshold. As a consequence, there is hardly
an enhancement at threshold as shown in Fig.~\ref{fig:top}. Moreover, individual initial
state radiation (ISR) and collective beamstrahlung effects further wash out the threshold.
Strong final state interaction effects cause a dependence
of the cross section at threshold on $\alpha_s$. Measuring an additional observable, however, like
the top momentum in addition to the cross section, allows to determine both $m_{\mathrm{t}}$ and 
$\alpha_s$ simultaneously. 

Top production is rather easily identified by requiring
one top quark to decay leptonically, $\mathrm{t \rightarrow W} q \rightarrow l \nu q$.
The momentum of the top quark can be determined from the other top quark that in 
general decays hadronically into three jets, $\mathrm{t \rightarrow W} q \rightarrow q q q$. 

\begin{figure}[htbp]
\begin{center}
\includegraphics*[width=0.50\linewidth]{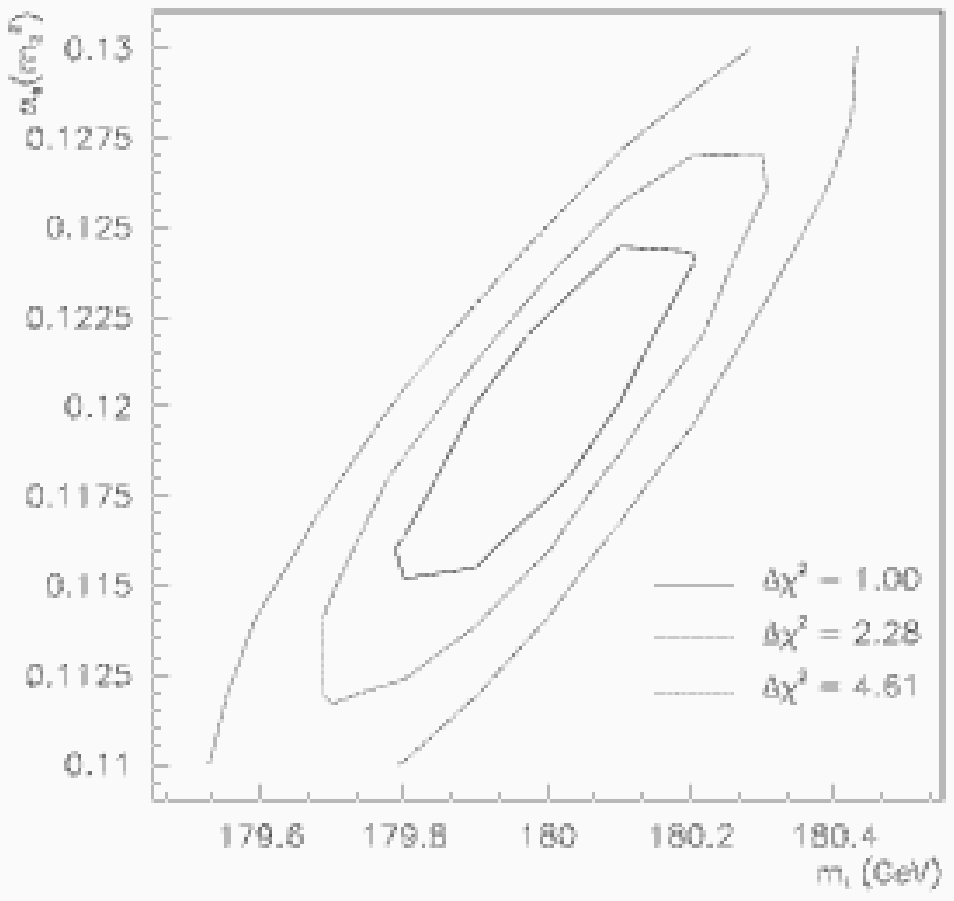} 
\includegraphics*[width=0.48\linewidth]{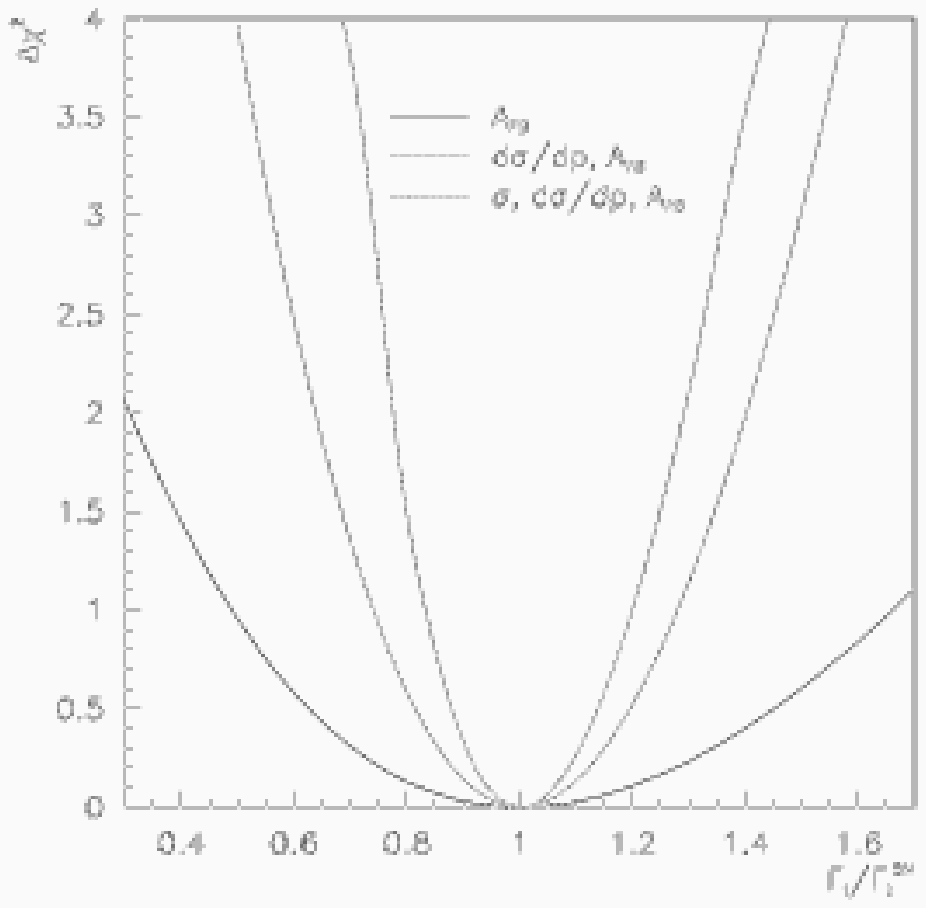} 
\end{center}
\caption{Left: Simultaneous determination of $m_{\mathrm{t}}$ and $\alpha_s$ in a threshold
scan. Right: Determination of the top width using the cross section, momentum distribution
and forward-backward asymmetry.\label{fig:mtas}}
\end{figure}

A simulated threshold scan has been evaluated\cite{Comas96}, with nine points equally spaced 
in energy, covering $356 < \sqrt{s} < 364$ GeV with a luminosity of 5 fb$^{-1}$ per point. 
Using both the cross section measurement and the top momentum distribution, one determines 
the top mass with an accuracy of about 200 MeV, as shown in Fig.~\ref{fig:mtas}. This will
improve the mass accuracy by a factor of ten with respect to LHC expectations.

Since the width of the loosely bound $\mathrm{t \bar{t}}$ systems at threshold is large,
interference between $S$ and $P$ states occurs and leads to a forward-backward asymmetry
in top production at threshold. While this asymmetry is rather insensitive to both 
$m_{\mathrm{t}}$ and $\alpha_s$, it gives an additional handle on the top width. 
Fig.~\ref{fig:mtas} shows that the top width can thus be determined to an accuracy of about 
20\%, using all observables from the threshold scan.

Continuum top production an the Linear Collider will allow all the usual measurements
of quark properties that have come from $\mathrm{e^+ e^-}$ machines for the lighter
quarks. The coupling of the top quark to the photon will tell about its
static electromagnetic properties and reveal an eventual form factor if it has substructure.
Likewise, its couplings to the Z will be measured. Its decay properties
will determine the $\mathrm{Wt\bar{b}}$ coupling. And, maybe most importantly, its coupling to
the Higgs boson will be measured, by observing Higgs decay into $\mathrm{t \bar{t}}$
if kinematically allowed, by virtual effects otherwise.

Experimental signatures of top quark production include very high particle multiplicities,
multi-jet and multi-lepton final states. Their efficient detection requires a high resolution,
fine granularity detector.

\section{Weak Boson Physics}\label{sec:weak}

The Higgs mechanism relates the masses of the weak bosons to their couplings 
in standard electroweak theory. At Born level one has
\beq 
 \sin^2{\theta_W} = 1 - M^2_{\mathrm{W}}/M^2_{\mathrm{Z}}
\eeq
where $\theta_W$ is the weak mixing angle defined by the ratio of weak charged and neutral
current couplings. Precision measurements of the couplings at LEP\cite{Mnich00} or in neutrino 
experiments\cite{Zeller99} can thus be expressed in terms of the W mass. The comparison
with direct measurements of the W mass at LEP and hadron colliders\cite{Mnich00}
provides one of the stringent tests of electroweak unification and the consistency of the 
Standard Model. 

The measurement of the Z boson mass is not likely to be improved over the precision reached at 
LEP in the foreseeable future. Even a dedicated run of Tesla at the Z pole\cite{Ohl99}, providing
$10^9$ Z events with polarised electrons and maybe also positrons will not improve on the 
mass precision which is limited by systematics. However, one may expect that the measurement
of $\sin^2{\bar{\theta}_W}$, the effective electroweak mixing angle, would improve by a 
factor of ten\cite{Moenig99}. The W mass measurement will also improve over current
errors, which are expected to reach 30 MeV at the end of LEP-200. LHC will improve this
measurement by a factor of about two. A one year dedicated run of the Linear Collider
at the threshold for $\mathrm{e^+ e^- \rightarrow W^+ W^-}$ at 161 GeV might
improve the mass error by another factor of two\cite{Ohl99}. Clearly,
nothing much else goes on at that particular energy and the measurement would be tedious.
It might eventually be worthwhile, if preceding measurements indicate a problem in this
sector.

\begin{figure}[htbp]
\begin{center}
\includegraphics*[width=\linewidth]{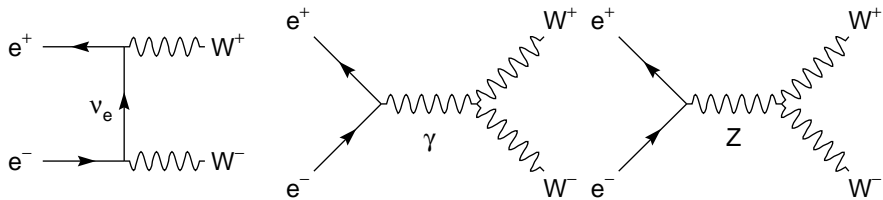} 
\end{center}
\caption{Born level Feynman graphs contributing to $\mathrm{e^+ e^- \rightarrow W^+ W^-}$.
\label{fig:eeww}}
\end{figure}

What will dramatically improve with a Linear Collider is the measurement of weak boson 
self couplings. These couplings define the electromagnetic and weak coupling constants via the
$\gamma \mathrm{W^+W^-}$ and $\mathrm{ZW^+W^-}$ vertices. They are measured
using the reaction $\mathrm{e^+ e^- \rightarrow W^+ W^-}$, as shown in Fig.~\ref{fig:eeww}.
The measurement would proceed in much the same way as at LEP now, using the total
cross section, the production angle and the decay angle distributions\cite{Mnich00}. 
Should deviations in these couplings be observed, they can be understood in terms of
an effective Lagrangian\cite{Hagiwara87} (imposing C  and P invariance for simplicity)
\bea
\frac{i{\cal L}_{\mbox{eff}}}{g_{VWW}} & = & 
 g_1^V V^\mu \left( W^-_{\mu\nu} W^{+\nu} - W^+_{\mu\nu} W^{-\nu} \right) \\
&   & \kappa_V W^+_\mu W^-_\nu V^{\mu\nu} +
\frac{\lambda_V}{M_W^2} V^{\mu\nu} W^{+\rho}_\nu W^-_{\rho\mu}
\eea
where the index $V=\gamma,\mathrm{Z}$ identifies the neutral boson, $V^\mu$ and $W^\mu$ are the
neutral and charged boson fields and $V^{\mu\nu}$ and $W^{\mu\nu}$ are their field tensors.
The Lagrangian is normalised such that $g_{\gamma WW} = -e$ and $g_{ZWW} = -e\cot{\theta_W}$,
so that the couplings $\kappa_V = 1$, $g_1^V = 1$ and $\lambda_V = 0$ reproduce the Standard
Model trilinear couplings. The photonic couplings relate to the static properties 
of the W boson, like its charge, $q_W = e g^\gamma_1$, its magnetic dipole moment
\beq
\mu_W = \frac{e}{2M_W}\left( g_1^\gamma+\kappa_\gamma+\lambda_\gamma \right) 
\eeq
and its electric quadrupole moment
\beq
{\cal Q}_W = -\frac{e}{M_W^2}\left( \kappa_\gamma-\lambda_\gamma \right).
\eeq
Electromagnetic gauge invariance requires $g_1^\gamma = 1$. SU(2)$\times$U(1) gauge
invariance requires that deviations of the couplings from their standard values respect
the relations
\bea
\Delta\kappa_Z & = & \Delta g^Z_1 - \Delta\kappa_\gamma\tan^2{\theta_W}  \\
\lambda_Z & = & \lambda_\gamma 
\eea
This way, the full wealth of fourteen anomalous trilinear couplings in the most general
Lagrangian is reduced to only three: $g_1^Z$, $\Delta\kappa_\gamma$ and $\lambda_\gamma$.
The existence of trilinear boson couplings is clearly established at LEP-200 and
anomalous couplings are limited to a few percent\cite{Mnich00}. 

At high energies, $\sqrt{s} - 2 M_{\mathrm{W}} \gg \Gamma_{\mathrm{W}}$, the structure of the
Lagrangian is the same for its $\gamma \mathrm{W^+W^-}$ and $\mathrm{ZW^+W^-}$ components.
Should anomalous couplings occur, it would thus be difficult to determine whether they are
of electromagnetic or weak origin. The two contributions can however be disentangled
with polarised beams, since the electromagnetic coupling is symmetric under $\mathrm{e}_L
\leftrightarrow \mathrm{e}_R$, while the weak coupling is asymmetric. In addition,
the so-called single W production reaction, 
$\mathrm{e^+ e^- \rightarrow e^- W^+ \bar{\nu}_e}$, which is dominated by $t$-channel 
$\gamma$W fusion, singles out the $\gamma \mathrm{W^+W^-}$ vertex. At $\sqrt{s} = 500$ GeV, with
an average electron polarisation of 80\% on unpolarised positrons, and using a luminosity
of 500 fb$^{-1}$, one can estimate the sensitivities\cite{Moenig99} to anomalous couplings 
as summarised in Tab.~\ref{tab:anoma}. An improvement of several orders of magnitude is 
achieved with respect to present knowledge. At even higher energy, $\sqrt{s} \simeq 1$ TeV, 
these results would even be 20 to 50\% better.

\begin{table}[htbp]
\caption{Statistical sensitivity to anomalous couplings expected from a collider run at 
$\sqrt{s} = 500$ GeV, with an average electron polarisation of 80\%, unpolarised positrons, 
and using a luminosity of 500 fb$^{-1}$.\label{tab:anoma}}
\begin{center}
\footnotesize
\begin{tabular}{|l|c|}
\hline
Parameter                      & Statistical Error $[\times 10^{-3}]$ \\ \hline
$\Delta \kappa_\gamma$         & 0.48 \\
$\lambda_\gamma$               & 0.72 \\ \hline
$\Delta g_1^{Z}$               & 2.50 \\
$\Delta \kappa_{Z}$            & 0.79 \\
$\lambda_{Z}$                  & 0.65 \\ \hline
\end{tabular}
\end{center}
\end{table}

\begin{figure}[htbp]
\begin{center}
\includegraphics*[width=0.75\linewidth]{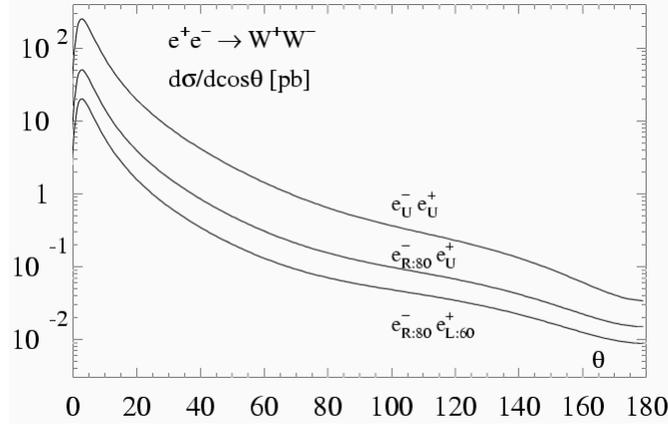} 
\end{center}
\caption{Born level differential cross section for $\mathrm{e^+ e^- \rightarrow W^+ W^-}$.
The unpolarised cross section is compared to the cross section with 80\%
electron beam polarisations as well as 80\% electron and 60\% positron beam polarisations.
\label{fig:wwpol}}
\end{figure}

Initial state polarisation can also be used to suppress large parts of the W pair production
cross section, if it is a background to other studies. This is demonstrated in 
Fig.~\ref{fig:wwpol}, where the unpolarised cross section is compared to 
cross sections with realistic
electron beam polarisations as well as electron and positron beam polarisations.

\section{Higgs Physics}\label{sec:higgs}

\begin{figure}[htbp]
\begin{center}
\includegraphics*[width=0.7\linewidth]{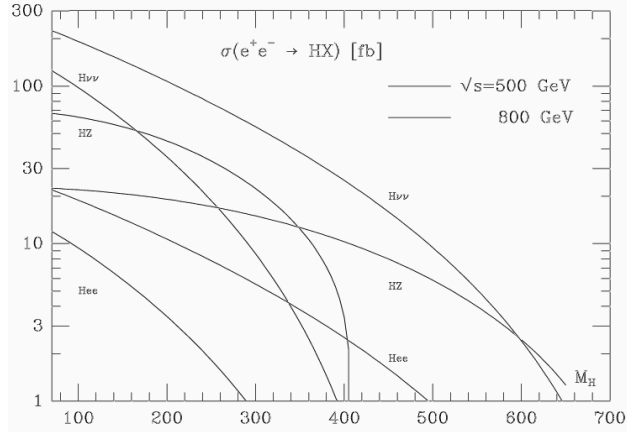} 
\end{center}
\caption{Higgs production cross section in the Standard Model 
as a function of the Higgs boson mass, at
$\sqrt{s} = 500$ GeV (solid lines) and $800$ GeV (dashed lines), separate for
the Higgsstrahlung process and the WW and ZZ fusion processes.
\label{fig:hxsec}}
\end{figure}

The Higgs boson, if it exists in either its Standard Model form or in
one of its  supersymmetric variants, will be discovered at LHC.  The
study of its properties is one of the strong arguments in favour of a
first generation Linear Collider. The production cross section for the
Standard Model Higgs boson is large, as shown in
Fig.~\ref{fig:hxsec}. For modest Higgs boson masses, the production is
dominated by boson fusion processes in $\mathrm{e^+ e^- \rightarrow H
e^+ e^-}$ and $\mathrm{e^+ e^- \rightarrow H \nu \bar{\nu}}$  final
states. The contribution of the Higgsstrahlung process  $\mathrm{e^+
e^- \rightarrow HZ}$, which has a sharper threshold, can be enhanced
relative to boson fusion by initial state polarisation.

\begin{figure}[htbp]
\begin{center}
\includegraphics*[width=\linewidth]{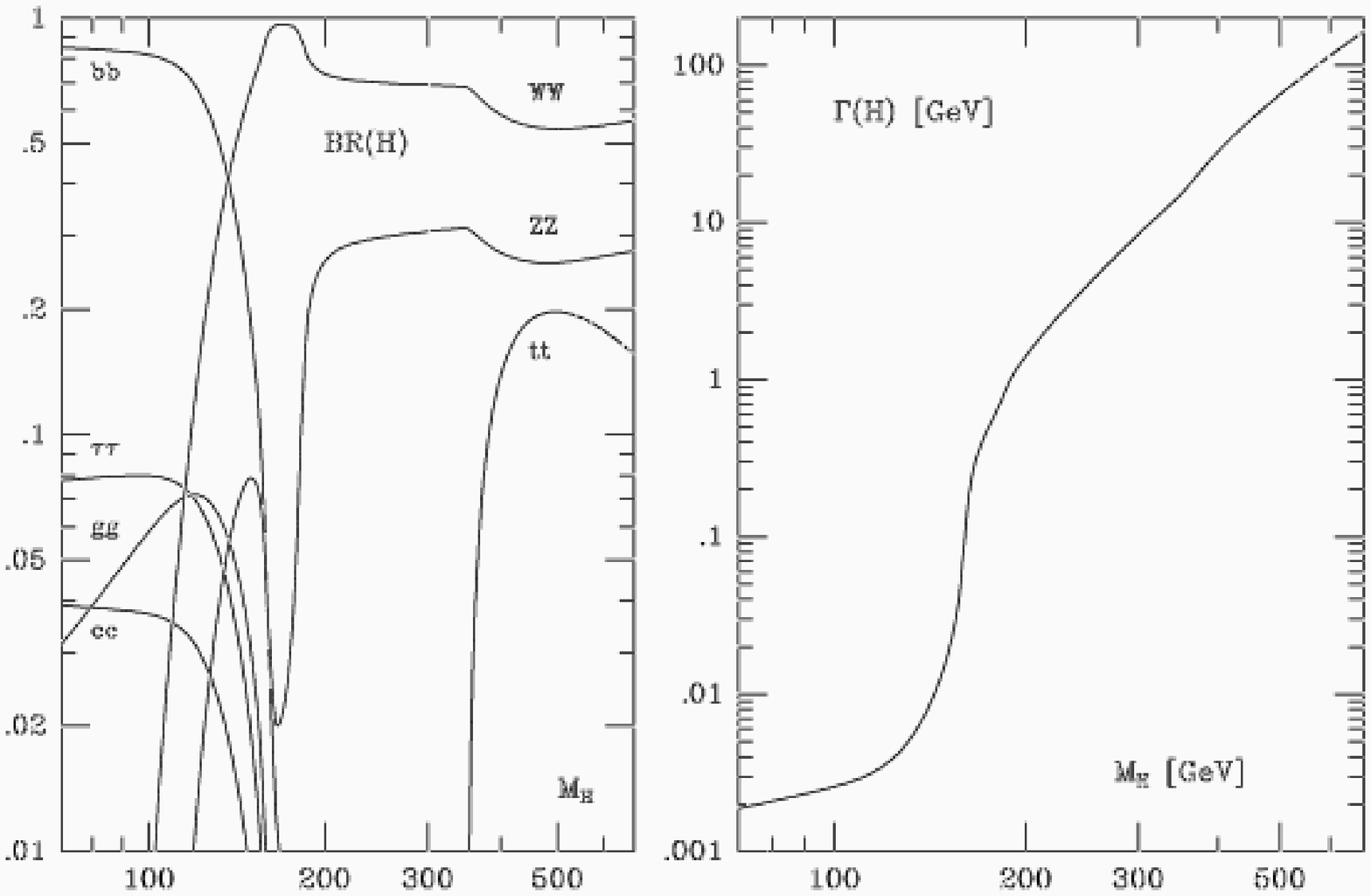} 
\end{center}
\caption{Higgs decay branching fractions (left) and total Higgs decay width (right)
as a function of the Higgs boson mass in the Standard Model.
\label{fig:hdecay}}
\end{figure}

The Higgs decay properties are completely defined once its mass, $M_{\mathrm{H}}$, is known.
Above the respective thresholds, one has
\bea
\Gamma(\mathrm{H} \rightarrow f \bar{f}) & = & 
\frac{G_F N_C}{4 \sqrt{2} \pi} M_f^2(M_{\mathrm{H}}^2) M_{\mathrm{H}} \\
\Gamma(\mathrm{H} \rightarrow V V) & = & N_V \frac{\sqrt{2} G_F}{32 \pi} M_{\mathrm{H}}
\eea
with the Fermi constant $G_F$, the number of fermion colours $N_C$, the constant $N_V=1,2$ 
for $V=\mathrm{Z,W}$ and the fermion and boson masses $M_f$ and $M_V$. The Higgs boson is 
thus a narrow state unless it is so heavy that it can decay into W or Z bosons. This is
demonstrated in Fig.~\ref{fig:hdecay}, which shows the branching fractions and the total
width as a function of the mass, for the Standard Model Higgs boson. 

A prime task for the Linear Collider will be to precisely measure the Higgs boson 
mass. A determination of the threshold position gives best results\cite{Wilson99}.
The mass error can be estimated as
\beq
\Delta M_{\mathrm{H}} = \sqrt{\sigma_{\mathrm{ZH}}} 
\left| \frac{\partial\!M_{\mathrm{H}}}{\partial\!\sigma_{\mathrm{ZH}}} \right|
\frac{1}{\sqrt{\epsilon P {\cal L}}}
\eeq
where $\sigma_{\mathrm{ZH}}$ is the Higgsstrahlung cross section, 
$\partial\!M_{\mathrm{H}}/\partial\!\sigma_{\mathrm{ZH}}$ is the inverse of
the cross section's sensitivity to mass, $\epsilon$ the experimental selection efficiency
for Higgs events, $P$ the average electron polarisation and ${\cal L}$ the
integrated luminosity. As an example, for a narrow Higgs of
$M_{\mathrm{H}} = 120$ GeV and ${\cal L} = 130$ fb$^{-1}$, one would have
$\Delta M_{\mathrm{H}} = 30 \mbox{MeV}/\sqrt{\epsilon P}$. For a 
wider one of 
$M_{\mathrm{H}} = 180$ GeV and ${\cal L} = 170$ fb$^{-1}$, the error would still be
$\Delta M_{\mathrm{H}} = 60 \mbox{MeV}/\sqrt{\epsilon P}$. 
With all Higgs decay channels observed, one can hope for $\epsilon P \simeq 0.5$, 
and thus a mass error of 50 to 100 MeV depending on the mass, as well as a Higgs 
width measurement with of order 10\% accuracy.

\begin{figure}[htbp]
\begin{center}
 \includegraphics*[width=0.7\linewidth]{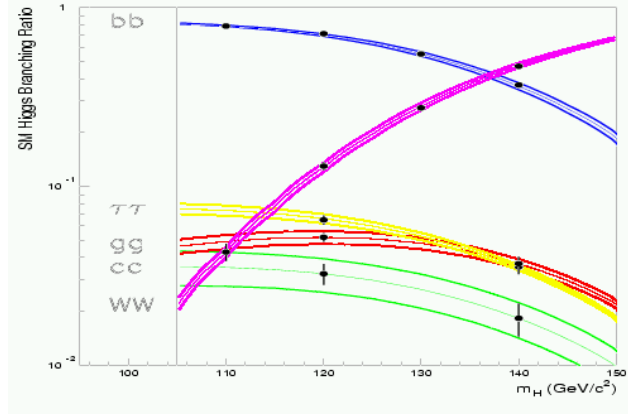}
\end{center}
\caption{Measurement of the Higgs decay branching fractions into fermion
and boson pairs.\label{fig:hbr}}
\end{figure}

The Higgs boson decay properties should establish that it indeed supplies the
functionality it has in the Standard Model, i.e. that it generates the masses of all
fermions and vector bosons. To verify that, one needs to measure its couplings to
all these particles and compare them to the prediction. It does not suffice to
cover only the dominant decay modes. In a very complete simulation study,
Battaglia\cite{Battaglia99} has shown that it is indeed possible to tag the majority
of Higgs decays using topological as well as secondary vertex tags fed into a 
neutral network. Decays into b and c as well as light quarks can thus be
separated from each other. This way, the decay branching fractions for
$\mathrm{H \rightarrow b \bar{b}}$, $\mathrm{c \bar{c}}$, $\tau^+ \tau^-$,
gluon and W pairs can be measured as shown in Fig~\ref{fig:hbr}. 
Using 100 fb$^{-1}$ of luminosity at 350 to 500 GeV center of mass
energy and for a Higgs boson mass between 110 and 140 GeV, one expects accuracies that 
vary between 5\% for $\mathrm{b \bar{b}}$ and 8\% for $\mathrm{c\bar{c}}$.

\begin{figure}[htbp]
\begin{center}
 \includegraphics*[width=0.7\linewidth]{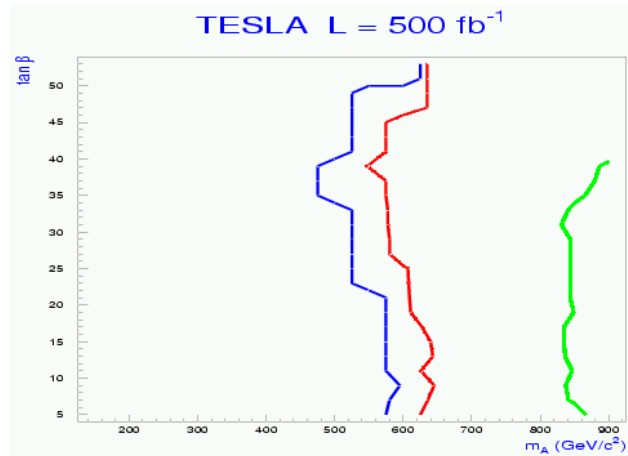}
\end{center}
\caption{Regions in the $M_{\mathrm{A}}$-$\tan{\beta}$ plane, 
where the MSSM is distinguishable from the Standard Model, given the Higgs branching
fractions measured at a Linear Collider. The lines correspond to 68\%, 90\% and
95\% confidence level from right to left, with the distinguishable regions to the left
of the lines.\label{fig:hmssm}}
\end{figure}

To judge this accuracy, a good way is to gauge it against close alternatives
to the Standard Model Higgs boson, like the lightest Higgs boson, $\mathrm{h}^0$, in minimal
supersymmetry (MSSM). The MSSM decay widths are approximately related to the
Standard Model ones by the supersymmetry parameters $\alpha$ and $\beta$
such that
\bea
\Gamma^{\mbox{MSSM}}_{\mathrm{b \bar{b}}} & \sim & \Gamma^{\mbox{SM}}_{\mathrm{b \bar{b}}} \frac{\sin^2 \alpha}{\cos^2 \beta}\\
\Gamma^{\mbox{MSSM}}_{\mathrm{c \bar{c}}} & \sim & \Gamma^{\mbox{SM}}_{\mathrm{c \bar{c}}} \frac{\cos^2 \alpha}{\sin^2 \beta}\\
\Gamma^{\mbox{MSSM}}_{\mathrm{W W^*}}     & \sim & \Gamma^{\mbox{SM}}_{\mathrm{W W^*}}
\eea
The parameter $\tan{\beta}$ is the ratio of the vacuum expectation values for the 
two Higgs boson doublets required in the MSSM. The other angle is defined
by $\tan{\alpha} = (M_{\mathrm{Z}}^2 - M_{\mathrm{A}}^2) 
\sin{\beta}\cos{\beta}/(M_{\mathrm{h}}^2 - 
M_{\mathrm{Z}}^2 \cos^2{\beta} - M_{\mathrm{A}}^2 \sin^2{\beta})$.   
With the branching ratio measurements alone one is thus able to distinguish
between the Standard Model Higgs boson and the lightest member of the MSSM
two doublet alternative in a large region of the MSSM parameter space,
as shown in Fig.~\ref{fig:hmssm}.

In addition to neutral Higgs bosons, supersymmetry requires the existence of charged Higgs
bosons, $\mathrm{H}^\pm$. These are copiously produced in $\mathrm{e^+ e^-}$ annihilation
and decay into fermion pairs like $\tau \bar{\nu}$, $\mathrm{b} \bar{\mathrm{t}}$ and
$\mathrm{s} \bar{\mathrm{c}}$, or boson pairs like $\mathrm{h W}$ or $\mathrm{A W}$.
These decays are characteristic enough to be selected with low backgrounds\cite{Accomando98}, 
and accurate mass measurements from the production threshold position can be envisaged.

\section{Supersymmetric Particles}\label{sec:susy}

Many theoretical arguments make it attractive to restore symmetry between matter made of
fermions and forces transmitted by vector bosons, by introducing 
supersymmetry\cite{Wess74}. It stabilises radiative corrections by cancellation 
between fermionic and bosonic loops\cite{Drees96}. It may also allow a grand unification of forces 
by introducing intermediate scales between the electroweak and the GUT 
scale\cite{Amaldi87}. However, introducing supersymmetry also opens Pandora's box in
that both the particle spectrum and the number of parameters in the theory expand
considerably. To make supersymmetric models at all predictive, additional assumptions
are necessary even for its minimal implementations. Popular models are based on minimal
supergravity\cite{Nilles84} (mSUGRA) or gauge mediated supersymmetry breaking\cite{Giudice99}
(GMSB). Both models have been studied extensively for the Linear 
Collider\cite{Martyn99,Ambrosanio99}.

\begin{figure}[htbp]
\begin{center}
 \includegraphics*[width=0.7\linewidth]{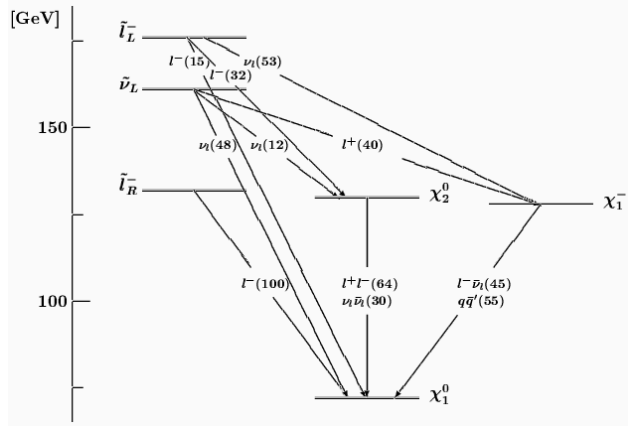}
\end{center}
\caption{Mass spectrum and decay modes of sleptons and light gauginos in a
mSUGRA scenario.\label{fig:susyspec}}
\end{figure}

What really matters at a Linear Collider is less the discovery of supersymmetric states.
Although a few of them like charginos and neutralinos are hard to produce 
at a hadron collider, one can assume that a fair fraction of the supersymmetric spectrum
will have been observed at the LHC. 
As an example, Fig.~\ref{fig:susyspec} shows a plausible spectrum of particles and decay modes
as expected in mSUGRA models. The r\^ole of the Linear Collider will
be to complete this spectrum and precisely determine masses and couplings. 
All this in order to tell which kind of supersymmetry
is realised in nature, by a determination of its (many) parameters.

\begin{figure}[htbp]
\begin{center}
\includegraphics*[width=0.5\linewidth]{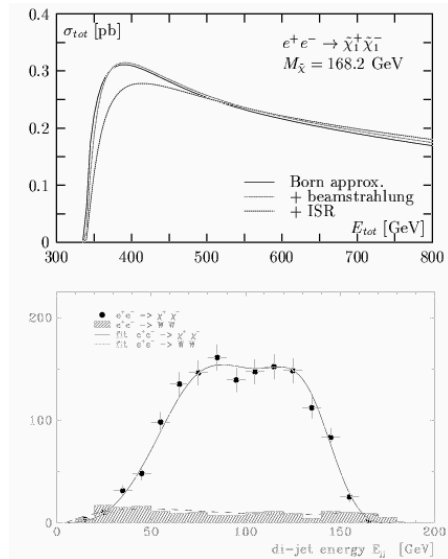} 
\end{center}
\caption{Top: Production cross section for chargino pairs of 168.2 GeV mass
as a function of center of mass energy. Bottom: Dijet energy spectrum in
continuum chargino decays at center of mass energies around 500 GeV. 
\label{fig:chi}}
\end{figure}

The chargino and neutralino production thresholds are fairly marked even when
ISR and beamstrahlung are taken into account. Fig.~\ref{fig:chi} shows
the production cross section and dijet invariant mass distributions for a chargino
mass of 170 GeV, at $\sqrt{s} \simeq 500$ GeV. One concludes that the mass of the  
lightest chargino can be measured with about 200 MeV accuracy from the continuum
production and 40 MeV from a threshold scan\cite{Martyn99}. Similar results are
obtained for neutralinos.

\begin{figure}[htbp]
\begin{center}
\includegraphics*[width=0.5\linewidth]{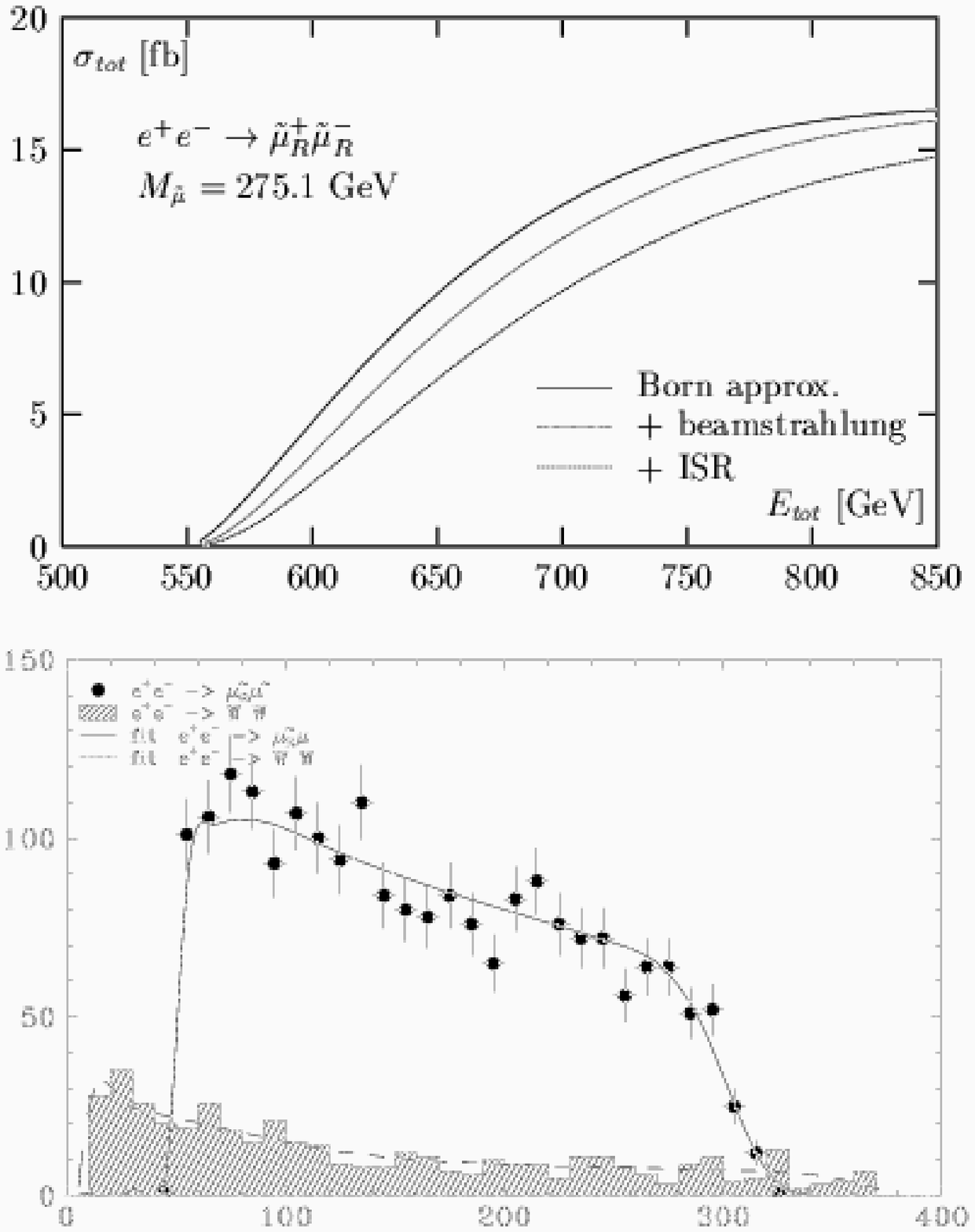} 
\end{center}
\caption{Top: Production cross section for smuon pairs of 275.1 GeV mass
as a function of center of mass energy. Bottom: Muon energy spectrum in
continuum smuon decays at center of mass energies around 800 GeV. 
\label{fig:smu}}
\end{figure}

The threshold for sleptons is smoother and the mass determination less accurate
as can be seen in Fig.~\ref{fig:smu}. Nevertheless the simultaneous study of
threshold cross section and continuum decay spectrum will allow the determination of 
slepton masses with a few hundred MeV accuracy\cite{Martyn99}.

These measurements can then be used in a RGE analysis of the mass spectrum to
determine the underlying SUSY model and its parameters.
Taking a mSUGRA scenario as a guideline example, one would thus expect to measure
the supersymmetry parameters with a respectable accuracy\cite{Martyn99}, as indicated in 
Tab.~\ref{tab:susy}. Polarisation of electrons and positrons is very important
for such a study since it allows to suppress dominant backgrounds.

\begin{table}[htbp]
\caption{Accuracy of supersymmetry parameters determined from the accessible mass 
spectrum of sleptons and light gauginos at a Linear Collider. As an example, 
parameters from a mSUGRA scenario have been used.\label{tab:susy}}
\begin{center}
\footnotesize
\begin{tabular}{|l|cc|}
\hline Parameter                & Assumed value  & Expected error
\\ \hline $m_0$                    & 100 GeV        & 0.09 GeV
\\ $M_1$                    & 200 GeV        & 0.20 GeV
\\ $M_2$                    & 200 GeV        & 0.20 GeV
\\ $A_0$                    & 0 GeV          & 10.3 GeV
\\ $\tan{\beta}$            & 3.0            & 0.04
\\ $sgn{(\mu)}$             & $+$            & fixed
\\ \hline
\end{tabular}
\end{center}
\end{table}

\section{Alternative New Physics}\label{sec:exotic}

\begin{figure}[htbp]
\begin{center}
\includegraphics*[width=0.69\linewidth]{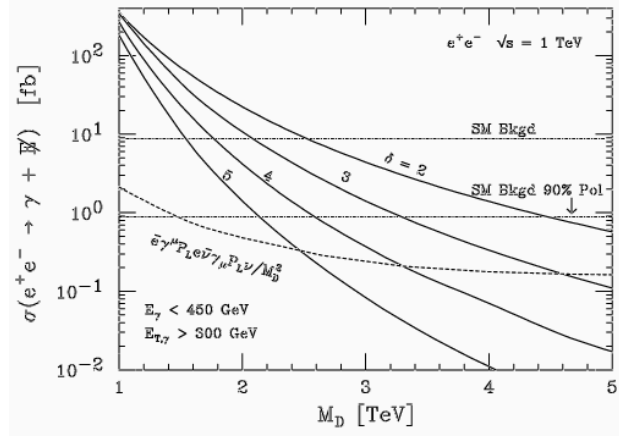}
\end{center}
\caption{Total cross section for the reaction $\mathrm{e^+ e^-} \rightarrow
\gamma +$ missing energy at $\sqrt{s} = 1$ TeV, as a function of the
fundamental mass scale $M_D$. Solid lines are for
graviton production with variable number of extra dimensions, $\delta$. 
The dashed-dotted lines represent Standard Model background without and
with 90\% electron beam polarisation.\label{fig:graviton}}
\end{figure}

In case nature chose not to make use of the Higgs mechanism, something
else must be responsible for particle masses. Candidates include extra
dimensions\cite{Giudice98}, which might also be transmitting
gravitational interactions\cite{Arkani98}. The mass of ordinary
particles might thus be an artefact of just observing propagation in a
subset of the existing dimensions, e.g.~via so-called Kaluza-Klein
towers. Gravitons might populate extra dimensions and gravity might be
of geometric origin after all. Effects of this idea at a Linear
Collider have been looked into\cite{Giudice98}. An example is given in
Fig.~\ref{fig:graviton}, which shows the cross section for one of the
signatures, single energetic photons, as a function of the fundamental
mass scale $M_D$. The signal would be an order of magnitude above
irreducible Standard Model background if the mass scale is not too
large.

\begin{figure}[htbp]
\begin{center}
\includegraphics*[width=0.5\linewidth]{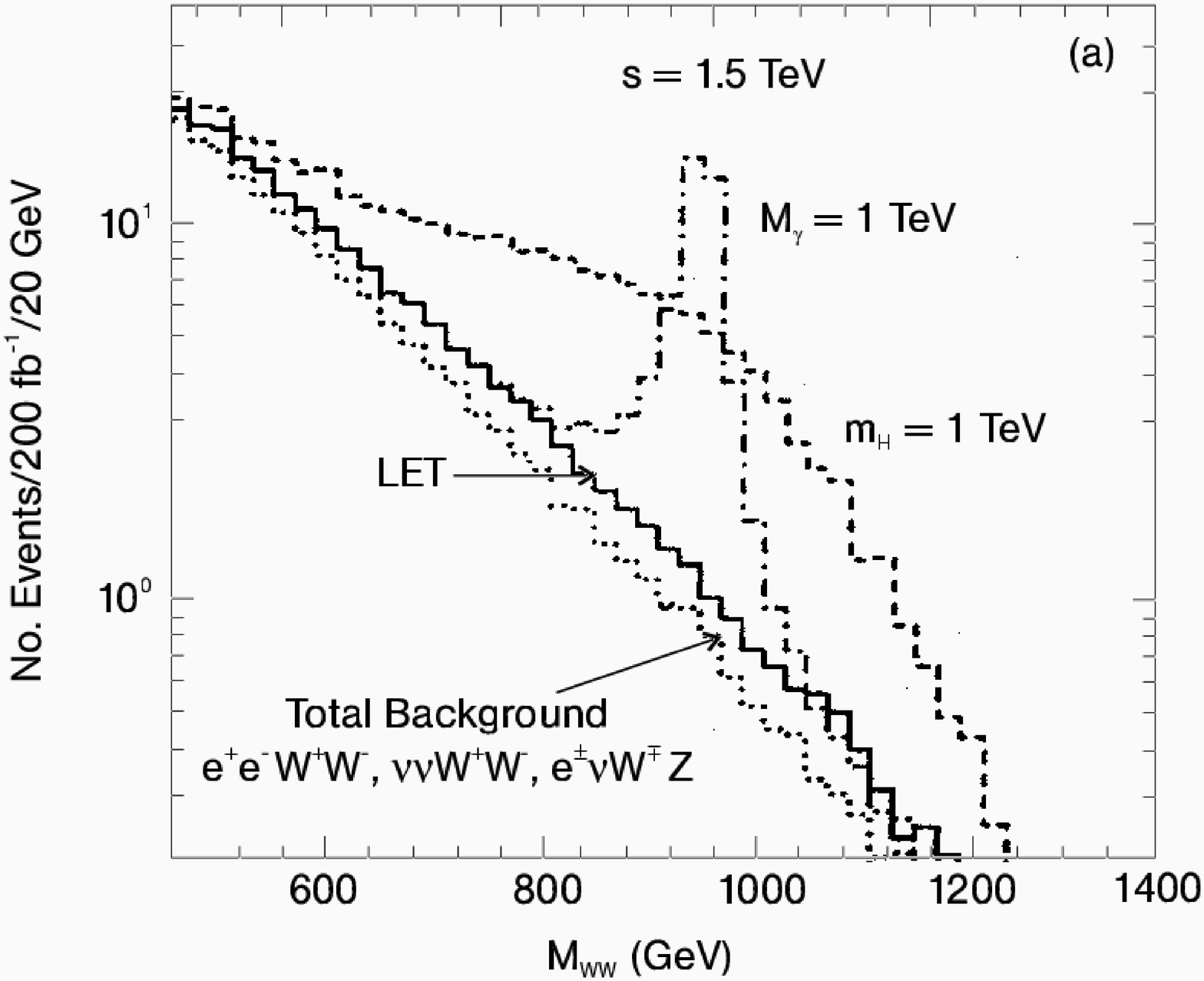}
\end{center}
\caption{Mass distribution of WW pairs for vector and scalar resonances
at a Linear Collider compared to Standard Model background.\label{fig:wwreso}}
\end{figure}

In the Standard Model, the Higgs particle also cancels divergences in
boson couplings, like WW scattering. If it does not exists, additional
strong inter-boson interactions might provide a cut off, instead. 
These would become apparent at a Linear Collider as vector or scalar
resonances between W bosons in $\mathrm{e^+ e^- W^+ W^-}$, 
$\mathrm{\nu \bar{\nu} W^+ W^-}$ and $\mathrm{\nu \bar{\nu} Z Z}$
final states\cite{Accomando98}, as shown in Fig.~\ref{fig:wwreso}. 

\begin{figure}[htbp]
\begin{center}
\includegraphics*[width=0.5\linewidth]{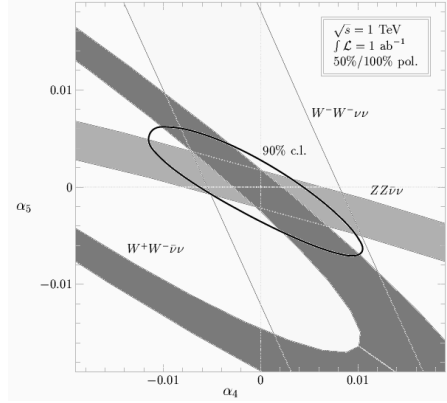}
\end{center}
\caption{Allowed regions in the expansion parameters in chiral
electroweak models from a cross section measurement for $\mathrm{W^+
W^-}$ and $\mathrm{Z Z}$ production at very high luminosities and beam
polarisations. Additional constraints might come from a measurement of
$\mathrm{e^- e^- \rightarrow \nu \nu W^- W^-}$.\label{fig:wwalpha}}
\end{figure}

From the cross section measurements, the couplings of the new
interactions are derived\cite{Boos98,Boos00}. As shown in Fig.~\ref{fig:wwalpha}, high
luminosity measurements at a Linear Collider cover the whole WW
threshold region and completely determine the relevant
couplings. Very high luminosities, of order 1000 fb$^{-1}$, and high beam polarisations are
required for such a measurement.

\begin{figure}[htbp]
\begin{center}
\includegraphics*[width=0.5\linewidth]{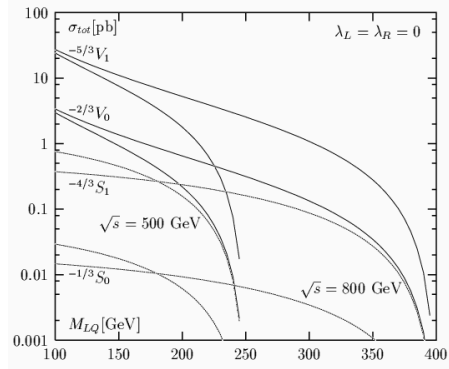} 
\end{center}
\caption{Total cross sections for scalar and vector leptoquark pair
production, at $\sqrt{s} = 800$ GeV, as a function of the leptoquark
mass. Corrections due to beamstrahlung and ISR are included.\label{fig:lq}}
\end{figure}

Since neither baryon nor lepton number is protected by a gauge
symmetry, nature might have states in stock that carry both quantum
numbers. The production cross section for leptoquarks at a Linear
Collider would be large, between 10 and 1800 fb$^{-1}$ depending on
the leptoquarks charge, spin and other
properties\cite{Accomando98,Ruckl99}, as shown in
Fig.~\ref{fig:lq}. Thus, if worst comes to worst, leptoquark
spectroscopy would be the {\em pi\`ece de r\'esistance} of
experimentation at a Linear Collider.

\section{Conclusions and Outlook}\label{sec:conclu}

Experiments at LEP have established the current effective theory, the
Standard Model. In these lectures, I have tried to substantiate my
view that there will be a sharing of responsibilities between LHC and a
future electron-positron Linear Collider. Experimentation at LHC will
cross the limits of validity of the Standard Model and outline a new
effective theory to replace it. A Linear Collider, covering {\em grosso
modo} the same energy range, is then needed to establish that next
effective theory. 

After almost 10 years of study, we can now be confident that the Tesla
project has a machine design that is both feasible and fulfils the
physics requirements for a first generation Linear Collider after
LHC. An excellent detector for Tesla can be built with reasonable
extrapolations of today's technologies.

Based on the Tesla Conceptual Design Report\cite{Brinkmann97},
optimisation of machine and detector is progressing rapidly. A
costed proposal will be published in the form of a Technical Design
Report later in 2000 or early in 2001. This will be followed by an in depth
discussion within the international particle physics community and with the
funding bodies, in the context of similar proposals elsewhere. In a
most optimistic scenario, if approved by 2003, first beams from Tesla
could be seen by 2008, at the earliest. 

Due to its built in limitations in center of mass energy, Tesla will
be followed by machines exploiting energy ranges beyond LHC. A very
promising contender would be a two beam accelerator like
CLIC\cite{clic00}, whose characteristics  might look like the summary
in Tab.~\ref{tab:clic}, and for which physics studies have just
started. Such a machine will clearly go beyond the 
regional scope of its predecessors. It will require a truly world-wide
effort, for the development of its novel acceleration technologies and
of the appropriate tools for experimentation. One might hope that
accelerator laboratories would form a world wide network to
support such a machine, following the example of experimental
collaborations that have done the same for a long time already.

\begin{table}[htbp]
\caption{Outlook to possible machine parameters at the CLIC two-beam collider 
project.\label{tab:clic}}
\begin{center}
\footnotesize
\begin{tabular}{|l|cc|}
\hline
                         & \multicolumn{2}{c|}{CLIC}                    \\ \hline
type                     & \multicolumn{2}{c|}{Two Beam Linear Collider}\\
maximum energy           & 1000 GeV       & 3000 GeV                    \\
total length             & 10km           & 40km                        \\
accelerating gradient    & 150 MV/m       & 170 MV/m                    \\
maximum bunches          &  \multicolumn{2}{c|}{154}                    \\
beam size at IP [$\mu$m] & 0.115$\times$0.002 & 0.030$\times$0.001      \\
luminosity [cm$^{-2}$s$^{-1}$] 
                         & 2$\times$10$^{34}$ & 10$\times$10$^{34}$     \\ \hline
\end{tabular}
\end{center}
\end{table}

\section*{Acknowledgements}

I would like to thank the organisers of this meeting, especially
Prof.~Manuel Aguilar Benitez, Dr.~Isabel Josa Mutuberr\'\i a and 
Dr.~Juan Alcaraz, for their fabulous hospitality, rooted in the Spanish
tradition. I would also like to thank all participants of the meeting
for creating a stimulating atmosphere of exchange and discussion.

\end{document}